\documentclass[conference]{IEEEtran}
\IEEEoverridecommandlockouts
\usepackage{graphicx}
\usepackage{amssymb}
\usepackage{verbatim}
\usepackage{amsmath}
\usepackage{multirow}
\usepackage{diagbox}
\usepackage{algorithm}
\usepackage[noend]{algpseudocode}
\usepackage{comment}  
\usepackage{moresize}
\newtheorem{def1}{\textbf{Definition}}
\newtheorem{def2}{\textbf{Lemma}}
\newtheorem{def3}{\textbf{Theorem}}
\newtheorem{def4}{\textbf{Note}}

\def\BibTeX{{\rm B\kern-.05em{\sc i\kern-.025em b}\kern-.08em
    T\kern-.1667em\lower.7ex\hbox{E}\kern-.125emX}}
\begin{document}

\title{Wireless Broadcast with short labels\\
}

\author{\IEEEauthorblockN{Gewu Bu\IEEEauthorrefmark{1}, Maria Potop-Butucaru\IEEEauthorrefmark{2}, Mika\"{e}l Rabie\IEEEauthorrefmark{3}}
\IEEEauthorblockA{Sorbonne University, LIP6, CNRS UMR 7606\\ Paris, France \\
Email: \IEEEauthorrefmark{1}gewu.bu@lip6.fr,
\IEEEauthorrefmark{2}maria.potop-butucaru@lip6.fr,
\IEEEauthorrefmark{3}mikael.rabie@lip6.fr}}

\maketitle

\begin{abstract}
In this paper, we study the broadcast problem in wireless networks when the broadcast is helped by a labelling scheme. We focus on two variants of broadcast: broadcast  without acknowledgment (i.e. the initiator of the broadcast  is not notified at the end of broadcast) and broadcast with acknowledgment. Our contribution is threefold. \emph{First}, we improve  in terms of memory complexity a recent \cite{ellen2019constant} labelling-based broadcast scheme with acknowledgment designed for arbitrary networks.
\emph{Second}, 
we  propose  label optimal broadcast algorithms in \emph{level separable networks} (a class of networks   issued from  recent studies in Wireless Body Area Networks).  
In this class of networks we propose an acknowledgment-free broadcast strategy using 1-bit labels and broadcast with acknowledgment using 2-bits labels. 
In the class of level-separable networks, our algorithms  finish within $2D$ rounds, where $D$ is the eccentricity of the broadcast  initiator. 
Interestingly, the time complexity of broadcast in the case of level-separable networks does not depend on the size of the network  but rather on  the initiator eccentricity which makes this class of graphs interesting for further investigation. 
\emph{Finally},  we study the hardness of determining that a graph  is  level separable. Our study shows that even though checking that a separation is  a level separation can be done in 
polynomial time, determining that a graph has the level separable property is NP-complete. This result opens interesting independent research directions.  

\end{abstract}


\section{Introduction}
\label{sec:in}
\emph{Broadcast} is the most studied communication primitive in networks and distributed systems. Broadcast ensures that  once a \emph{source node} (a.k.a. the broadcast initiator) sends a message, all other nodes in the network should receive this message in a finite time. Limited by the transmission range, messages might not be sent directly from one node to some other node in the network. Therefore relay nodes need to assist the source node during the message propagation by re-propagating it. Deterministic centralized broadcast, where nodes have  complete network knowledge, has been studied by Kowalski {\it et al.}  in \cite{kowalski2007optimal}. The authors propose an optimal solution  that completes within $O(D \log^2n)$ rounds, where $n$ is the number of nodes in the network and $D$ is the largest distance from the source to any node of the network. 
For deterministic distributed broadcast, assuming that nodes only know their IDs (i.e. they do not know the  IDs of their neighbours nor the  network topology), in \cite{cicalese2009faster} is proposed the fastest broadcast within $O(n\log D \log \log D)$ rounds, where $D$ is the diameter of the network. The lower bound in this case, proposed in \cite{clementi2001selective}, is $\Omega(n\log D)$.

In wireless networks, when a message is sent from a  node it goes into the wireless channel in the form of a wireless signal which may be received by all the nodes within the transmission range of  the sender node. However, when a node is located in the range of more than one node that sends messages simultaneously, the multiple wireless signals may generate \emph{collisions}  at the receiver. The receiver  cannot decode any useful information from the superimposed interference signals.
At the MAC layer, several solutions have been proposed in the last two decades in order to reduce collisions. All of them offer probabilistic guarantees.  
Our study follows the recent work that addresses this problem at the application layer. More specifically, we are interested in deterministic solutions for broadcasting messages based on the use of extra information or advise (also referred to as \emph{labelling}) precomputed before the broadcast invocation.   

Labelling schemes have been designed to compute network size, the father-son relationship and the geographic distance between arbitrary nodes  in the network  (e.g. \cite{abiteboul2001compact}, \cite{gavoille2004distance} and \cite{gorain2018finding}).  Labelling schemes have been also used  in \cite{fraigniaud2010local} and \cite{glacet2017time} in order to improve the efficiency of Minimum Spanning Tree or Leader Election algorithms. Furthermore, \cite{cohen2008label} and \cite{fraigniaud2008tree}  exploit labelling in order to improve the existing solutions for network exploration by a robot/agent moving in the network. 

Very few works ( e.g. \cite{ilcinkas2010fast} and \cite{ellen2019constant}) exploit labelling schemes to design efficient \emph{broadcast} primitives. When using labelling schemes, nodes record less information than in the case of centralized broadcast, where nodes need to know complete network information. Compared with the existing solutions for deterministic distributed broadcast  the time complexity is improved.
In \cite{ilcinkas2010fast} the authors prove that for an arbitrary network, to achieve broadcast within a constant number of rounds, a $O(n)$ bits of advice is sufficient but not $o(n)$. Very recently, a labelling scheme with  2-bits advice (3 bits for broadcast with acknowledgment)  is proposed in \cite{ellen2019constant}. The authors prove that their algorithms need $2n-3$ rounds for the broadcast without acknowledgment and $3n-4$  rounds for broadcast with acknowledgment in an arbitrary network.  

\emph{Contribution:} 
Our work is in the line of research described in  \cite{ellen2019constant}. We first  improve  in terms of  memory complexity  the broadcast scheme with acknowledgment proposed in \cite{ellen2019constant}. Differently, from the solution proposed in \cite{ellen2019constant}, our solution does not use extra local persistent memory except the 3-bits labels. Then,  we study labelling-based broadcast in a new family of networks, called level-separable networks issued from Wireless Body Area Networks (e.g.  \cite{badreddine2015broadcast}, \cite{badreddine2017peak}, \cite{bu2017total}, \cite{badreddine2017convergecast} and \cite{bu2018ban}). 
In this class of networks we propose an acknowledgment-free broadcast strategy using 1-bit labels and a broadcast scheme with acknowledgment using 2-bits labels. 
Our algorithms terminate within $2D$ rounds for both types of broadcast primitives, where $D$ is the eccentricity of the  broadcast source. Interestingly, the time complexity  of broadcast in the case of level separable networks 
does not directly depend on  the network size which makes the study of level separable networks of independent interest.
We further investigate the hardness of determining if a graph  is  level separable. Our study shows that even though checking that a separation is  a level separation can be done in 
polynomial time, determining that a graph has the level separable property is NP-complete. This result opens interesting independent research directions that will be discussed in the conclusion of this document.  
  
\section{Model and Problem Definition}
We model the network as a \emph{graph} $G=(V, E)$ where $V$, the set of \emph{vertices}, represents the set of nodes in the network and $E$, the set of $edges$, is a set of unordered pairs $e=(u, v)$, $u, v \in V$, that represents the communications links between nodes $u$ and $v$. In the following $d(u)$ denotes the set of neighbours of node $u$.
%
We assume that the network is connected, i.e., there is a path between any two nodes in the network.

We assume that nodes execute the same algorithm and are \emph{time synchronized}. The system execution is decomposed in $rounds$.  When a node $u$ sends a message at round $x$, all nodes in $d(u)$ receive the message at the end of round $x$.  Collisions occur at node $u$ in round $x$ if a set of nodes, $M \subseteq d(u)$ and $|M| > 1$, send a message in round $x$. In that case, it is considered that $u$ has not received any message.

In the following we are interested in solving the \emph{Broadcast problem}: when a source node $s$ sends a data message $\mu$, this $\mu$ should be received by all the nodes in the network in a finite bounded  time. We are also interested in solving  
\emph{Broadcast with acknowledgment problem}: once all nodes received $\mu$ , an acknowledgment message, called $ACK$, will be generated and sent backward to the source node $s$ in a finite bounded time.


\section{Broadcast with ACK for arbitrary networks}
\label{descACK}
In \cite{ellen2019constant} the authors propose a broadcast with acknowledgment algorithm $\beta_{ACK}$ for general networks using a 3-bits labelling scheme $\lambda_{ACK}$ . 
The idea of the broadcast algorithm $\beta_{ACK}$ is an extension  of algorithm $\beta$ also described  in \cite{ellen2019constant} which implements the broadcast of a message $\mu$ within bounded time. 
At each round, only nodes that received  
$\mu$ in specified previous rounds can send it to avoid the potential collisions. Initially, the source node $s$ sends $\mu$ to all its neighbours. A \emph{Frontier Set}, $Frnt$, is defined  where $Frnt$ contains all nodes that have not received $\mu$  and that 
have direct connections with nodes received $\mu$ at the end of that round. Then a \emph{Minimal Dominating Set}, $miniD$ is defined  over the nodes that already have received $\mu$ such that nodes in $Frnt$ are dominated by nodes in $miniD$. Nodes in $miniD$ then send  $\mu$, so that some of nodes in $Frnt$ can receive $\mu$.   $Frnt$ and $miniD$ are therefore updated since some nodes will leave $Frnt$ and may join  $miniD$ in the next round. Nodes in new $miniD$ will continue send $\mu$ until $Frnt = \varnothing$. The broadcast then finishes. Note that during the execution, a node in $miniD$ at round $i$ may stay in the $miniD$ till round $j$, where $i<j$. In this case, additional notification message $Stay$ is needed to be sent to nodes who need to stay in $miniD$. 

Algorithm $\beta_{ACK}$ extends $\beta$ by adding an additional ACK message, that is, when the last nodes receive $\mu$, one of them will generate an $ACK$ message that will be forwarded back to $s$. During the execution, nodes will store the round number at which they received and sent $\mu$ with two variables $informedRound$ and $transmitRounds$. So that nodes know which path the $ACK$ should follow back to $s$. $\beta_{ACK}$ is based on a 3 bits  labelling scheme $\lambda_{ACK}$. The first bit, $X_1$, indicates if a node $u$ will be in $miniD$ at least once during the broadcast. If yes, then $X_1$ of $u$ equals  $1$; if not, it equals   $0$. If $X_1$ of $u$ equals   $1$, when $u$ receives message $\mu$, $u$ can re-send it once. The second bit $X_2$ of $u$ equal to $1$ means that $u$ needs to send a $Stay$ when it receives $\mu$ to notify the sender of $\mu$ to stay in $miniD$ for the next round. Only one of the informed nodes will have the third bit $X_3$ equal to $1$. This node will generate the $ACK$ to be sent back to $s$.
At the end of the broadcasting, which finishes in $2n-3$ rounds, the last informed node generates and sends back to the source node  the $ACK$ within additional $n-2$ rounds, where $n$ is the number of nodes in the network. 

Our optimization with respect to the $\lambda_{ACK}$ proposed  in \cite{ellen2019constant} comes from the following simple observation:
in a 3-bits labelling, there are 8 possible states: 000, 001, 010, 011, 100, 101, 110 and 111. The algorithm in \cite{ellen2019constant}  uses only 5 of them: 000, 001, 010, 100 and 110. In this section, we propose a labelling scheme, $\lambda_{oACK}$ and a broadcast scheme with $ACK$ algorithm that use  all the 8 states of the 3-bits labelling in order to improve the memory complexity of  the solution proposed in  \cite{ellen2019constant}.
The idea of our optimization is as follows: instead of only using the last bit $X_3$ (the third bit) as a marker to point who is (one of) the last informed node(s) during the broadcast, we use also this third bit to show  a path back to the source node $s$ from the last informed node. Differently, from the solution proposed in  \cite{ellen2019constant}, nodes do not need to keep additional variables  in order to send back the $ACK$ during the execution. Our proposition can therefore, save node's memory and computational power.

In the following, we present our $\lambda_{oACK}$ labelling scheme and $\beta_{oACK}$ algorithm.

\subsection{3-bits Labelling Scheme $\lambda_{oACK}$} 
\label{OACK}
The first two bits of the labelling scheme $X_1$ and $X_2$  have the same functionality as in the  $\lambda_{ACK}$ scheme of \cite{ellen2019constant}. 
The intuitive idea is as follows: 1) $X_1 = 1$ for nodes  who should propagate $\mu$ when they receive it; 2) $X_2 = 1$ for nodes that need to send $Stay$ back to their sender neighbour to notice that they need to stay in $miniD$ and send $\mu$ one more time in the next round; 3) $X_3 = 1$ for one of the last receiving nodes to generate $ACK$ and send it back to the source node $s$. In our scheme $\lambda_{oACK}$ we also set $X_3$ (the third bit) to 1 for all nodes on the  path back from the last informed node (who holds $001$) to $s$. Note that, nodes on that path could have four kinds of different labels: $101$, $011$, $111$ and $001$, where $001$ is the label of the last informed node. Label states $101$, $011$ and $111$ are not used in the original $\beta_{ACK}$, therefore nodes can easily recognize if they are on the path  to transmit $ACK$ back to $s$.
Note that we do not change the main architecture of the algorithm $\beta_{ACK}$ with labelling scheme $\lambda$ proposed in \cite{ellen2019constant}, therefore the correctness proof of our algorithm is very similar to the one in \cite{ellen2019constant}. See Section \ref{A1} for a detailed proof that follows the lines of the proof in \cite{ellen2019constant}.

\subsection{Broadcast Algorithm $\beta_{oACK}$}

\begin{algorithm} [t] \footnotesize
\caption{$\beta_{oACK}(\mu)$ executed at each node $v$}
\label{alg2}
\begin{algorithmic}
\State \%Each node has a variable $sourcemsg$. The source node has this variable initially set to $\mu$, all other nodes have it initially set to $null$. 
\For{each round $r$ from 0}
\If{$v$ is source node and $r = 0$}
\State transmit $sourcemsg$
\EndIf
\If{$v$ is not source node}
\If{message $m$ is received AND m $\neq$ "stay"} 
\State $sourcemsg \gets m$
\EndIf
\ElsIf{The node received $\mu$ before round $r$}
\If{$v$ received $sourcemsg$ for first time in round $r-2$}
\If{$X_1 =  1$}
\State transmit $sourcemsg$
\EndIf
\ElsIf{$v$ received $sourcemsg$ for first time in round $r-1$}
\If{$X_1 = 0$ and $X_2 = 0$ and $X_3 =  1$}
\State transmit "ACK"
\ElsIf{$X_2 = 1$}
\State transmit "stay"
\EndIf
\ElsIf{$v$ received "stay" in round $r-1$}
\If{$v$ transmitted $sourcemsg$ in round $r-2$}
\State transmit $sourcemsg$
\EndIf
\ElsIf{$v$ received "ACK" in round $r-1$}
\If{$X_3 =  1$}
\State transmit "ACK"
\EndIf
\EndIf
\EndIf
\EndFor
\end{algorithmic}
\end{algorithm}

Our broadcast algorithm $\beta_{oACK}$ that uses $\lambda_{oACK}$ described above is described as Algorithm \ref{alg2}.
Nodes  with $X_1 = 1$ receiving a message at round $i-1$ send  it at round $i$. Then nodes who send at round $i$ wait for the $stay$ message, at round $i+1$, from nodes with $X_2 = 1$. If nodes who send at round $i$ receive a $stay$ at round $x+1$, they continue to send  one more time $\mu$ at round $i+2$, otherwise, they stay silent. When nodes with label $001$ receive the message, they generate an $ACK$ and send it. Since $\lambda_{oACK}$ already marked the path back   to the source node, in Algorithm $\beta_{oACK}$, the $ACK$ message will only be re-propagated by nodes with $X_3 = 1$. i.e., nodes with label $101$, $111$ and $011$.

Note that our proposed Algorithm $\beta_{oACK}$ does not need additional variables to reconstruct the path back to $s$ during the broadcast execution. In  Algorithm $\beta_{ACK}$ \cite{ellen2019constant}, two additional variables $informedRound$ (type $int$) and $transmitRounds$ (type \emph{table of int}) are needed to rebuild the back-way path. $informedRound$ is used to record the round number in which a node received $\mu$; $transmitRounds$ is a table used to record all the round numbers in which one node transmits $\mu$. However, by using $\beta_{oACK}$, the $ACK$  transmission processing can be completed only by checking the third bit, $X_3$. Our Algorithm $\beta_{oACK}$ does not need any extra local storage for detecting the path for $ACK$.

\subsection{Correctness of $\beta_{oACK}$}
\label{A1}
Our proposition of $\beta_{oACK}$ with $\lambda_{oACK}$ is based on the algorithm $\beta$ with labelling scheme $\lambda$ proposed in \cite{ellen2019constant}. The algorithm $\beta_{oACK}$ can be seen as the combination of  two phases: \emph{Broadcast Phase} and \emph{ACK Phase}. The aim of the broadcast phase of $\beta_{oACK}$ is to finish first the broadcast:  every node in the network should be informed of the message $\mu$ sent by the source node. 
In the second phase, one of the last informed nodes will generate $ACK$ and send it back to the source node through a specific path marked according to the labelling scheme $\lambda_{oACK}$.

These two phases are well separated, because $ACK$ will only be generated and sent to the network after one of the last informed nodes received $\mu$ sent by the source node. Therefore, there will be no collisions between $\mu$ and $ACK$ during the execution of $\lambda_{oACK}$. 

During the first broadcast phase, we use exactly the same idea of the algorithm $\beta$ with labelling scheme $\lambda$ in \cite{ellen2019constant}. The correctness of this phase is given as Theorem \ref{Tellen2019constant} in \cite{ellen2019constant}, as follows:

\begin{def3} \cite{ellen2019constant}
\label{Tellen2019constant}
Consider any $n$-node unlabelled graph G with a designated source node $s$ with $\mu$. By applying the 2-bits labelling scheme $\lambda$ and then executing algorithm $beta$, all nodes in $G \setminus \{s\}$ are informed within $2n - 3$ rounds. 
\end{def3}

As described in Section \ref{descACK}, 
the idea is that   every two rounds, if there are still nodes that have not received $\mu$ yet, a non-null subset of these nodes will form the new $Frnt$. When the new $miniD$ set of nodes send $\mu$, some nodes belonging to $Frnt$ will receive it. Then the number of the non-informed nodes will decrease until $0$. In the worst case, when the topology of the network is a line, $\mu$ has to go through all of them one by one to reach every node. The algorithm therefore finishes within $2n$ rounds.

We then prove that the ACK phase of $\beta_{oACK}$,   finishes within $n$ rounds.

\begin{def2}
\label{lemappACK}
After the broadcast phase finishes during the execution of $\beta_{oACK}$,  $ACK$ will be sent back to the source node within $n$ rounds.
\end{def2}

\textbf{Proof} 
By using $\lambda_{oACK}$ described in Section \ref{OACK}, only one of the last informed nodes $u$ will have its three bits equal to $001$. Then $u$ will send $ACK$, and only nodes with $X_3 = 1$ can forward $ACK$ back only when they received it. The back-forward path to the source node is chosen by $\lambda_{oACK}$. In the worst case, when the topology of the network is a line, then the $ACK$ has to go through all the nodes to reach the source node. Therefore, during the execution of $\beta_{oACK}$, $ACK$ will need at most  $n$ rounds to reach the source node.

The Theorem \ref{Tellen2019constant} and Lemma \ref{lemappACK} therefore complete the correctness proof of $\beta_{oACK}$.




\section{Level-Separable Networks}
\label{defLVS}
In this section, we define a family of networks,  \emph{Level-Separable Network}, issued from WBAN (Wireless Body Area Networks) area (e.g. \cite{DBLP:journals/corr/abs-1303-2062}, \cite{Latre:2011:SWB}, \cite{WBAN1}, \cite{badreddine2015broadcast}, \cite{badreddine2017peak}, \cite{bu2017total}, \cite{badreddine2017convergecast} and \cite{bu2018ban}), then we investigate the broadcast problem in these networks.

\subsection{Motivation for the study of level-separable networks}
The motivation of the study of level-separable networks comes from the recent studies of WBAN.
 WBAN is similar to WSN (Wireless Sensor Networks) in terms of devices functionalities and architecture. However, WBAN still has important differences with WSN. The deployment environment and application scenario make them  totally different: WSN is usually deployed in  wide range  areas; WBAN on the other hand, is deployed on (or inside) the human body, to detect various physiological parameters of the human body. WBAN devices are in close contact with the body, therefore the transmission power cannot has a setting as high as in the case of WSN. Using a relatively small transmission power in WBAN might be greatly affected by the absorption, interference and refraction of the human body.

Furthermore, WBAN has to face the challenge of the human body mobility, which makes the connexion between nodes appear and disappear from time to time. The challenge in WBAN is how to improve the communication reliability of the network by taking into  consideration  the  human  mobility  and the changes in the  communication channels.

To our best knowledge, Naganawa {\em et. al} \cite{naganawa2015simulation} proposed the first simulation-based \emph{Data Sets} of the human mobility and the channel quality change. These data sets provide measurement results of channel attenuation between different WBAN devices deployed on different positions of the human body during different human movement actions. The data sets have been validated by comparing to massive real-human based measurement results.

\begin{figure}[!ht]
\centering
\includegraphics[width=0.5\textwidth]{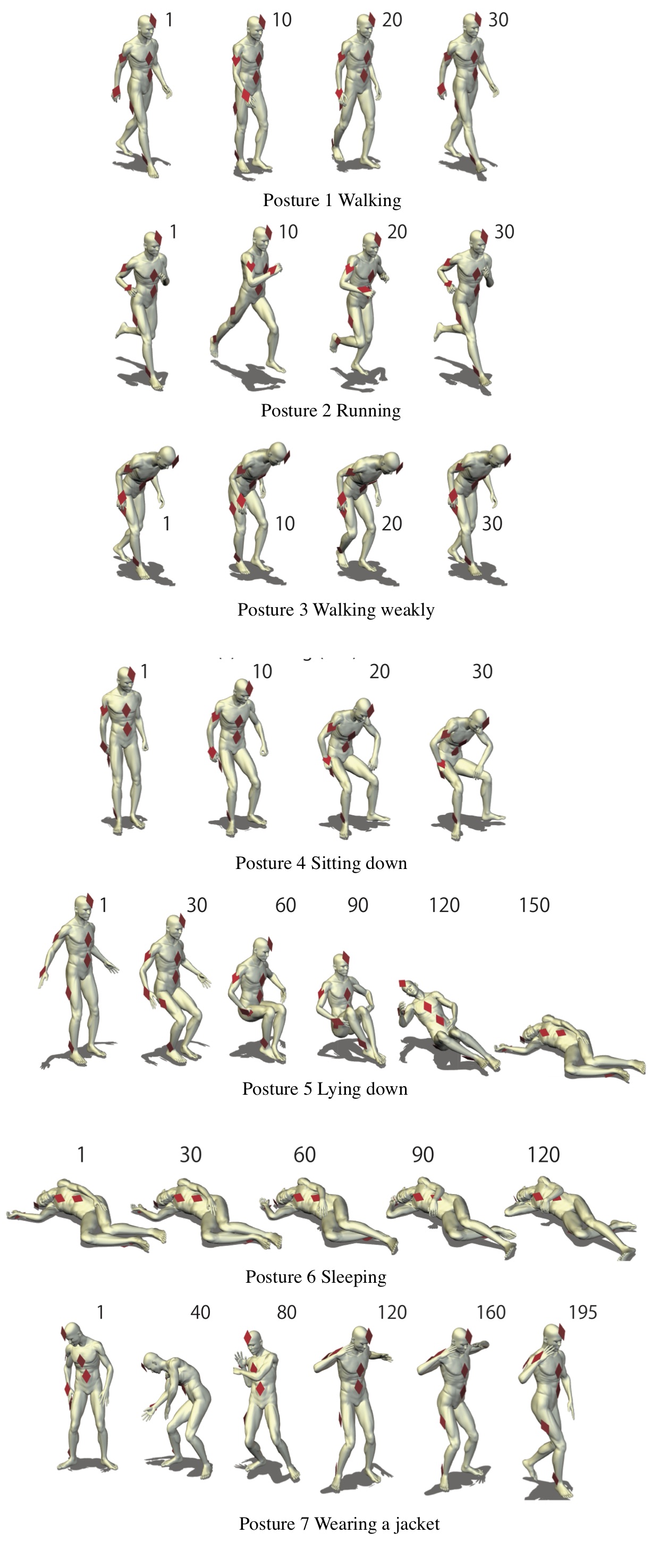}
\caption{7 Different Human Postures \cite{naganawa2015simulation}}
\label{walk1}
\end{figure}

The network architecture of proposed environment is composed of seven WBAN devices distributed on the body as follows:
Navel, Chest, Head, Upper Arm, Ankle, Thigh and Wrist. The authors measure the connectivity between every two nodes in seven different postures: 1)Walking, 2)Running, 3)Walking weakly, 4)Sitting down, 5)Lying down, 6)Sleeping and 7)Putting on a jacket, respectively  (see Figure \ref{walk1}).

In each posture, a continuous human action has been decomposed into a set of frames. Each single human body picture with a corresponding frame number, $x$, is a screenshot of this continuous human action at the $x$th frame. For example, in posture 1)Walking (see Figure \ref{walk1}), the continuous action takes 30 frame, and it uses four screenshots at 1st frame, 10th frame, 20th frame and the 30th frame, respectively to represent this action. The red diamonds in the figures represent sensors on the human body while the body is moving. 

Tables \ref{dataSet1} to \ref{dataSet7} show the measurement results of channel attenuation between two nodes pair in seven different human mobility postures. Values above the main diagonal represent the mean values of the random channel attenuation between any two WBAN nodes of the body. Based on data sets from \cite{naganawa2015simulation}, authors of \cite{badreddine2015broadcast} propose a channel-mobility model: for every wireless signal sent from a WBAN node, a random attenuation is added to the outgoing communication channel. If the signal strength after the attenuation is smaller than the sensitivity of the receiver, it will be dropped. The random attenuation is calculated by different normal distributions specified by means and standard deviations for each couple of nodes (e.g, the random channel attenuation between nodes on head and on upper arm in posture 1)Walking has the mean 45.4dB and the standard deviations 5.1dB).


\begin{table}[!ht]
\centering
\caption{Means and Standards Deviations of Path Loss for all the links in Posture 1) Walking \cite{naganawa2015simulation}}
\label{dataSet1}
\resizebox{.5\textwidth}{!}{
\begin{tabular}{c|ccccccc|c}
\hline
 \huge{$T_{X}$ or $R_{X}$} & \huge{navel} & \huge{chest} & \huge{head} & \huge{upper arm} & \huge{ankle} & \huge{thigh} & \huge{wrist} & \\ \hline
\huge{navel} & \diagbox[width=6em,trim=l] & \huge{30.6} & \huge{45.1} & \huge{44.4} & \huge{57.4} & \huge{45.8} & \huge{41.0} &  \\
\huge{chest} & \huge{0.5} & \diagbox[width=6em,trim=l] & \huge{38.5} & \huge{40.6} & \huge{58.2} & \huge{51.6} & \huge{45.1} &  \\
\huge{head} & \huge{0.8} & \huge{0.5} & \diagbox[width=6em,trim=l] & \huge{45.4} & \huge{64.0} & \huge{61.3} & \huge{49.7} &  \\
\huge{upper arm} & \huge{5.8} & \huge{5.2} & \huge{5.1} & \diagbox[width=6em,trim=l] & \huge{54.2} & \huge{45.5} & \huge{34.0} & \huge{Mean[dB]} \\
\huge{ankle} & \huge{4.3} & \huge{3.4} & \huge{5.0} & \huge{3.1} & \diagbox[width=6em,trim=l] & \huge{40.6} & \huge{48.9} &  \\
\huge{thigh} & \huge{2.0} & \huge{2.5} & \huge{6.8} & \huge{4.8} & \huge{1.0} & \diagbox[width=6em,trim=l] & \huge{35.0} &  \\
\huge{wrist} & \huge{5.0} & \huge{3.6} & \huge{3.8} & \huge{2.5} & \huge{3.8} & \huge{3.3} & \diagbox[width=6em,trim=l] &  \\ \hline
 &  &  &  \multicolumn{3}{c}{\huge{Standard deviation [dB]}}  &  &  & 
\end{tabular}
}
\end{table}

\begin{table}[!ht]
\centering
\caption{Means and Standards Deviations of Path Loss for all the links in Posture 2) Running \cite{naganawa2015simulation}}
\label{dataSet2}
\resizebox{.5\textwidth}{!}{
\begin{tabular}{c|ccccccc|c}
\hline
 \huge{$T_{X}$ or $R_{X}$} & \huge{navel} & \huge{chest} & \huge{head} & \huge{upper arm} & \huge{ankle} & \huge{thigh} & \huge{wrist} & \\ \hline
\huge{navel} & \diagbox[width=6em,trim=l] & \huge{31.4} & \huge{47.4} & \huge{54.5} & \huge{57.9} & \huge{44.8} & \huge{45.9} &  \\
\huge{chest} & \huge{1.4} & \diagbox[width=6em,trim=l] & \huge{41.0} & \huge{39.2} & \huge{61.0} & \huge{49.9} & \huge{41.2} &  \\
\huge{head} & \huge{3.5} & \huge{2.9} & \diagbox[width=6em,trim=l] & \huge{41.3} & \huge{65.6} & \huge{59.3} & \huge{45.5} &  \\
\huge{upper arm} & \huge{9.9} & \huge{8.4} & \huge{8.4} & \diagbox[width=6em,trim=l] & \huge{58.0} & \huge{52.4} & \huge{33.8} & \huge{Mean[dB]} \\
\huge{ankle} & \huge{6.9} & \huge{6.9} & \huge{5.7} & \huge{8.2} & \diagbox[width=6em,trim=l] & \huge{39.0} & \huge{56.9} &  \\
\huge{thigh} & \huge{2.0} & \huge{2.5} & \huge{6.8} & \huge{4.8} & \huge{1.0} & \diagbox[width=6em,trim=l] & \huge{49.6} &  \\
\huge{wrist} & \huge{6.1} & \huge{8.2} & \huge{3.5} & \huge{4.6} & \huge{7.5} & \huge{11.6} & \diagbox[width=6em,trim=l] &  \\ \hline
 &  &  &  \multicolumn{3}{c}{\huge{Standard deviation [dB]}}  &  &  & 
\end{tabular}
}
\end{table}


\begin{table}[!ht]
\centering
\caption{Means and Standards Deviations of Path Loss for all the links in Posture 3) Walking weakly \cite{naganawa2015simulation}}
\label{dataSet3}
\resizebox{.5\textwidth}{!}{
\begin{tabular}{c|ccccccc|c}
\hline
 \huge{$T_{X}$ or $R_{X}$} & \huge{navel} & \huge{chest} & \huge{head} & \huge{upper arm} & \huge{ankle} & \huge{thigh} & \huge{wrist} & \\ \hline
\huge{navel} & \diagbox[width=6em,trim=l] & \huge{26.1} & \huge{42.4} & \huge{44.3} & \huge{55.4} & \huge{44.9} & \huge{34.0} &  \\
\huge{chest} & \huge{0.4} & \diagbox[width=6em,trim=l] & \huge{38.1} & \huge{37.3} & \huge{58.8} & \huge{47.1} & \huge{41.7} &  \\
\huge{head} & \huge{1.3} & \huge{0.7} & \diagbox[width=6em,trim=l] & \huge{44.5} & \huge{52.4} & \huge{60.0} & \huge{42.8} &  \\
\huge{upper arm} & \huge{5.5} & \huge{5.5} & \huge{6.8} & \diagbox[width=6em,trim=l] & \huge{53.7} & \huge{45.1} & \huge{34.5} & \huge{Mean[dB]} \\
\huge{ankle} & \huge{4.2} & \huge{4.6} & \huge{3.3} & \huge{6.1} & \diagbox[width=6em,trim=l] & \huge{42.4} & \huge{49.2} &  \\
\huge{thigh} & \huge{2.2} & \huge{5.3} & \huge{5.4} & \huge{4.8} & \huge{2.2} & \diagbox[width=6em,trim=l] & \huge{37.9} &  \\
\huge{wrist} & \huge{2.8} & \huge{2.5} & \huge{1.5} & \huge{3.1} & \huge{4.8} & \huge{4.4} & \diagbox[width=6em,trim=l] &  \\ \hline
 &  &  &  \multicolumn{3}{c}{\huge{Standard deviation [dB]}}  &  &  & 
\end{tabular}
}
\end{table}


\begin{table}[!ht]
\centering
\caption{Means and Standards Deviations of Path Loss for all the links in Posture 4) Sitting down \cite{naganawa2015simulation}}
\label{dataSet4}
\resizebox{.5\textwidth}{!}{
\begin{tabular}{c|ccccccc|c}
\hline
 \huge{$T_{X}$ or $R_{X}$} & \huge{navel} & \huge{chest} & \huge{head} & \huge{upper arm} & \huge{ankle} & \huge{thigh} & \huge{wrist} & \\ \hline
\huge{navel} & \diagbox[width=6em,trim=l] & \huge{27.9} & \huge{41.1} & \huge{41.5} & \huge{59.6} & \huge{48.3} & \huge{38.6} &  \\
\huge{chest} & \huge{1.0} & \diagbox[width=6em,trim=l] & \huge{37.0} & \huge{36.0} & \huge{60.0} & \huge{51.0} & \huge{43.2} &  \\
\huge{head} & \huge{1.6} & \huge{0.8} & \diagbox[width=6em,trim=l] & \huge{42.1} & \huge{63.7} & \huge{59.1} & \huge{46.9} &  \\
\huge{upper arm} & \huge{5.3} & \huge{4.8} & \huge{6.3} & \diagbox[width=6em,trim=l] & \huge{63.7} & \huge{49.0} & \huge{37.7} & \huge{Mean[dB]} \\
\huge{ankle} & \huge{8.4} & \huge{8.0} & \huge{8.7} & \huge{8.1} & \diagbox[width=6em,trim=l] & \huge{40.9} & \huge{60.2} &  \\
\huge{thigh} & \huge{6.3} & \huge{5.3} & \huge{7.8} & \huge{5.5} & \huge{6.3} & \diagbox[width=6em,trim=l] & \huge{35.1} &  \\
\huge{wrist} & \huge{4.6} & \huge{5.3} & \huge{5.5} & \huge{5.7} & \huge{9.6} & \huge{6.9} & \diagbox[width=6em,trim=l] &  \\ \hline
 &  &  &  \multicolumn{3}{c}{\huge{Standard deviation [dB]}}  &  &  & 
\end{tabular}
}
\end{table}


\begin{table}[!ht]
\centering
\caption{Means and Standards Deviations of Path Loss for all the links in Posture 5) Lying down \cite{naganawa2015simulation}}
\label{dataSet5}
\resizebox{.5\textwidth}{!}{
\begin{tabular}{c|ccccccc|c}
\hline
 \huge{$T_{X}$ or $R_{X}$} & \huge{navel} & \huge{chest} & \huge{head} & \huge{upper arm} & \huge{ankle} & \huge{thigh} & \huge{wrist} & \\ \hline
\huge{navel} & \diagbox[width=6em,trim=l] & \huge{30.5} & \huge{45.1} & \huge{54.1} & \huge{65.0} & \huge{55.8} & \huge{49.7} &  \\
\huge{chest} & \huge{2.2} & \diagbox[width=6em,trim=l] & \huge{38.2} & \huge{43.4} & \huge{63.6} & \huge{54.3} & \huge{46.5} &  \\
\huge{head} & \huge{3.3} & \huge{1.3} & \diagbox[width=6em,trim=l] & \huge{40.0} & \huge{61.8} & \huge{58.6} & \huge{45.5} &  \\
\huge{upper arm} & \huge{5.9} & \huge{4.2} & \huge{4.2} & \diagbox[width=6em,trim=l] & \huge{58.3} & \huge{50.1} & \huge{38.8} & \huge{Mean[dB]} \\
\huge{ankle} & \huge{6.9} & \huge{5.8} & \huge{7.0} & \huge{5.1} & \diagbox[width=6em,trim=l] & \huge{41.2} & \huge{44.7} &  \\
\huge{thigh} & \huge{12.4} & \huge{10.1} & \huge{10.1} & \huge{10.1} & \huge{7.2} & \diagbox[width=6em,trim=l] & \huge{41.6} &  \\
\huge{wrist} & \huge{6.3} & \huge{4.9} & \huge{3.8} & \huge{1.9} & \huge{9.6} & \huge{8.8} & \diagbox[width=6em,trim=l] &  \\ \hline
 &  &  &  \multicolumn{3}{c}{\huge{Standard deviation [dB]}}  &  &  & 
\end{tabular}
}
\end{table}


\begin{table}[!ht]
\centering
\caption{Means and Standards Deviations of Path Loss for all the links in Posture 6) Sleeping \cite{naganawa2015simulation}}
\label{dataSet6}
\resizebox{.5\textwidth}{!}{
\begin{tabular}{c|ccccccc|c}
\hline
 \huge{$T_{X}$ or $R_{X}$} & \huge{navel} & \huge{chest} & \huge{head} & \huge{upper arm} & \huge{ankle} & \huge{thigh} & \huge{wrist} & \\ \hline
\huge{navel} & \diagbox[width=6em,trim=l] & \huge{31.7} & \huge{64.3} & \huge{66.5} & \huge{72.5} & \huge{56.3} & \huge{58.6} &  \\
\huge{chest} & \huge{4.3} & \diagbox[width=6em,trim=l] & \huge{50.9} & \huge{51.9} & \huge{72.4} & \huge{51.3} & \huge{44.1} &  \\
\huge{head} & \huge{10.4} & \huge{10.6} & \diagbox[width=6em,trim=l] & \huge{39.0} & \huge{69.4} & \huge{59.9} & \huge{42.5} &  \\
\huge{upper arm} & \huge{4.6} & \huge{2.7} & \huge{11.3} & \diagbox[width=6em,trim=l] & \huge{51.5} & \huge{42.7} & \huge{30.9} & \huge{Mean[dB]} \\
\huge{ankle} & \huge{5.7} & \huge{7.5} & \huge{9.3} & \huge{0.8} & \diagbox[width=6em,trim=l] & \huge{35.7} & \huge{56.8} &  \\
\huge{thigh} & \huge{5.0} & \huge{2.1} & \huge{10.8} & \huge{2.6} & \huge{0.9} & \diagbox[width=6em,trim=l] & \huge{48.9} &  \\
\huge{wrist} & \huge{7.8} & \huge{4.1} & \huge{7.2} & \huge{3.6} & \huge{2.8} & \huge{2.5} & \diagbox[width=6em,trim=l] &  \\ \hline
 &  &  &  \multicolumn{3}{c}{\huge{Standard deviation [dB]}}  &  &  & 
\end{tabular}
}
\end{table}


\begin{table}[!ht]
\centering
\caption{Means and Standards Deviations of Path Loss for all the links in Posture 6) Wearing a jack \cite{naganawa2015simulation}}
\label{dataSet7}
\resizebox{.5\textwidth}{!}{
\begin{tabular}{c|ccccccc|c}
\hline
 \huge{$T_{X}$ or $R_{X}$} & \huge{navel} & \huge{chest} & \huge{head} & \huge{upper arm} & \huge{ankle} & \huge{thigh} & \huge{wrist} & \\ \hline
\huge{navel} & \diagbox[width=6em,trim=l] & \huge{27.4} & \huge{43.3} & \huge{56.8} & \huge{62.8} & \huge{45.0} & \huge{52.0} &  \\
\huge{chest} & \huge{3.4} & \diagbox[width=6em,trim=l] & \huge{37.4} & \huge{51.4} & \huge{60.4} & \huge{47.7} & \huge{50.9} &  \\
\huge{head} & \huge{4.9} & \huge{3.6} & \diagbox[width=6em,trim=l] & \huge{49.2} & \huge{64.0} & \huge{51.7} & \huge{46.8} &  \\
\huge{upper arm} & \huge{6.7} & \huge{5.1} & \huge{9.2} & \diagbox[width=6em,trim=l] & \huge{52.3} & \huge{52.9} & \huge{31.1} & \huge{Mean[dB]} \\
\huge{ankle} & \huge{7.1} & \huge{9.9} & \huge{8.8} & \huge{4.1} & \diagbox[width=6em,trim=l] & \huge{39.5} & \huge{55.1} &  \\
\huge{thigh} & \huge{2.5} & \huge{6.3} & \huge{7.0} & \huge{5.1} & \huge{1.7} & \diagbox[width=6em,trim=l] & \huge{52.3} &  \\
\huge{wrist} & \huge{7.4} & \huge{5.9} & \huge{5.9} & \huge{4.8} & \huge{10.8} & \huge{7.7} & \diagbox[width=6em,trim=l] &  \\ \hline
 &  &  &  \multicolumn{3}{c}{\huge{Standard deviation [dB]}}  &  &  & 
\end{tabular}
}
\end{table}

\begin{figure}[t]
\centering
\includegraphics[width=0.5\textwidth]{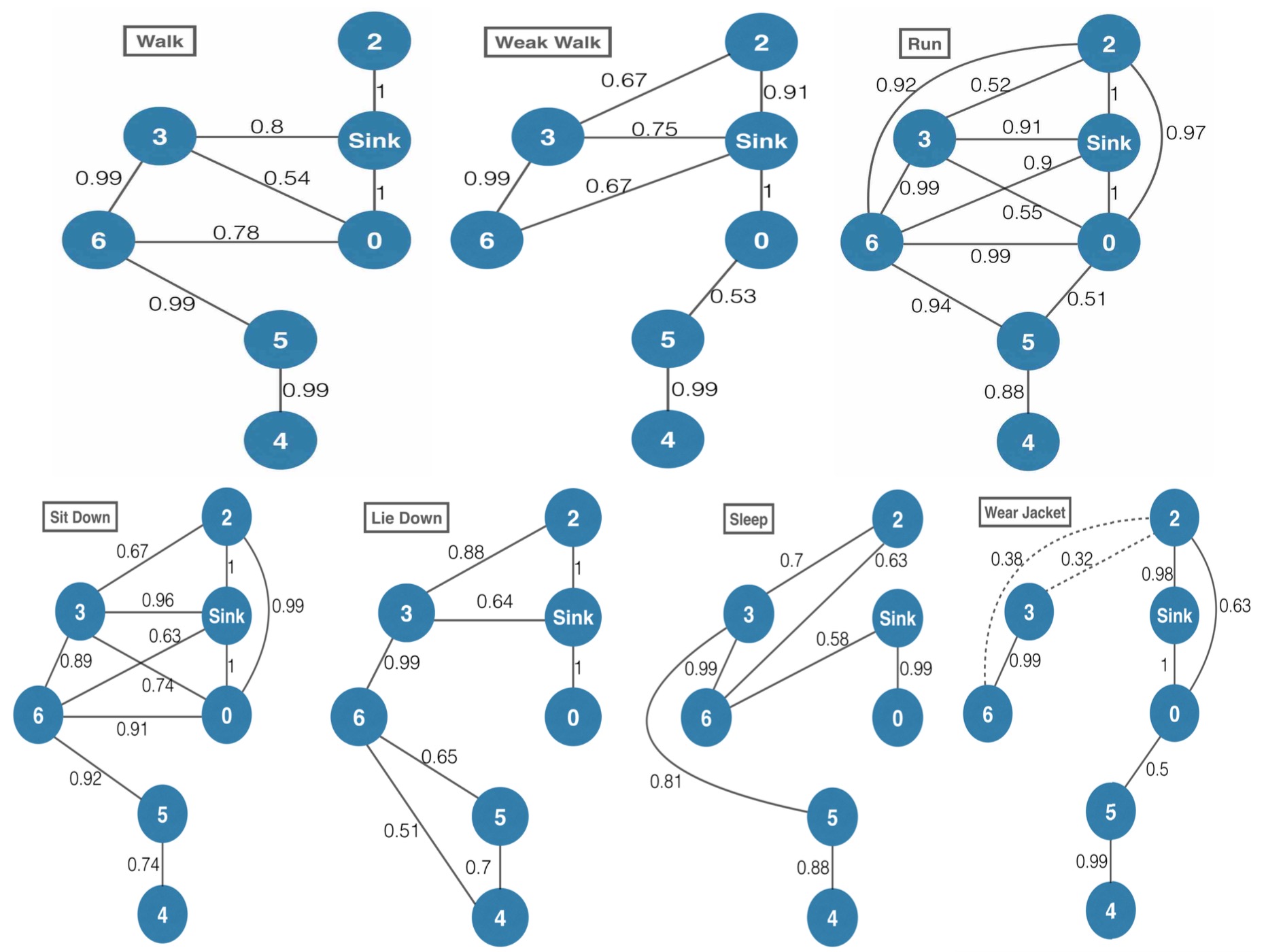}
\caption{Graphs that model human postures in WBAN. Numbers on the edges represent the edge reliability \cite{badreddine2017peak}}
\label{realworld}
\end{figure}

Studies (\cite{bu2017total} and \cite{badreddine2017peak}) conducted in WBAN show that various postural mobilities can be modeled  as graphs (one for each human posture),  see Figure \ref{realworld}. Moreover, the authors in \cite{makovmodel} proved that the performances of any protocol for wireless body area networks strongly depend on the topology of the graph and it should be noted that none of the graphs corresponds to the classical classes (e.g. planar or minor-free).  
 
 In the case presented above (the only available to date benchmark for practical WBAN),  each graph is a level-separable network defined below.
 


\subsection{Formal definition of level-separable networks}

We say that an arbitrary network is a Level-Separable Network if the underlay communication graph $G = (V, E)$ of the network verifies the \emph{Level-Separable} property defined below. To define the \emph{Level-Separable} property, we introduce some preliminary notations.

Let  $G(V, E)$ be a network and let $s \in V$, a predefined vertex, be the source node of the broadcast.  Each vertex $u \in V$ has a geometric distance with respect to $s$ denoted $d(s, u)$. The eccentricity of vertex $s$,  $\varepsilon_G(s)$, is the farthest distance from $s$ to any other vertex. In the rest of the paper, we denote $\varepsilon_G(s)$ by $D$. 

\begin{def1}[Level]
 Let $G(V,E)$ be a network and $s$ the source node. For any vertex $u$ in $G(V,E)$, the $level$ of  $u$ is $l(u) = d(s, u)$ is its  geometric distance to $s$. Let 
$S_i = \{u \mid u \in V, \ l(u) =i \}$
denote the set containing all the vertices at level $i$. 
\end{def1}

\begin{def1}[Parents and Sons]
Let $G(V,E)$ be a network.
A vertex $u$ is   parent of   vertex $v$ (a vertex $v$ is  son of   vertex $u$) in graph $G$ with the root source node  $s$: if 
$l(v) - l(u) = 1 \ \wedge \{u,v\} \in E$.
Let $S(u)$ ($P(v)$) be the set of sons (parents) of $u$ ($v$). If $v \in S(u) $ ($u \in P(v)$), we say that $u$ ($v$) has $v$ ($u$) as  son (parent). 
\end{def1}

\emph{Level-Separable} property below defines how to filter nodes in the same level $i$ into two disjoint subsets.

\begin{def1}[Level-Separable Subsets]
Given $G(V, E)$ a network and the set $S_i$ (the set of all vertices in the same level $i$ of $G$), the level-separable subsets of $S_i$ are $S_{i,1}$ and $S_{i,2}$, such that 
$S_{i,1} \cap S_{i,2} = \varnothing, \ S_{i,1} \cup S_{i,2} = S_i$
\end{def1}

There may be many possible pairs of  $S_{i,1}$ and $S_{i,2}$ for a level $i$. Let $T_i$ be the set of all possible pairs of \emph{Level-Separable Subsets}:
\[
T_i = \{(S_{i,1}^{(1)}, S_{i,2}^{(1)}), (S_{i,1}^{(2)}, S_{i,2}^{(2)}),..., (S_{i,1}^{(2^x)}, S_{i,2}^{(2^x)})\}
\]
 
where $(m)$ on right-top of each pair represent the index of pairs (the $m$th pairs) in $T_i$, and $x = |S_i|$.



\begin{def1}[Level-Separable Property]
\label{def5}
Given an arbitrary graph $G(V,E)$, for all level $i \in [1, D-1]$, where $D$ is the eccentricity of the source node, $G$ verifies the Level-Separable property, if there are pairs for every $T_i$, $(S_{i,1}^{(k)}, S_{i,2}^{(k)})$, such that:
$\forall u\in S_{i+1},|P(u) \cap S_{i,1}^{(k)}| = 1\vee |P(u) \cap S_{i,2}^{(k)}| = 1$
i.e., for every vertex $u$ at level $i+1$, $u$ has only one parent in $S_{i,1}$ or $S_{i,2}$. 
\end{def1}
Note that when $S_{i,1}$ is fixed, $S_{i,2}$ is $S_i \setminus S_{i,1}$. 
 
\begin{def1}[Level-Separable Network]
A network  $G(V,E)$ is a Level-Separable Network if its underlay graph verifies the Level-Separable property. 
\end{def1}
Note that \emph{Level-Separable Graph} has a similar flavour with \emph{Bipartite Graph} \cite{gross2005graph}. A graph $G=(V,E)$ is said to be Bipartite if and only if there exists a partition $V=A \cup B$ and $A \cap B= \varnothing$. So that all edges share a vertex from both sets $A$ and $B$, and there is no edge containing two vertices in the same set. A bipartite graph  separates nodes   into two independent sets. In a level-separable network, we aim at  separating nodes of the  same level. Moreover, we  are interested  in the relation between the two separated  sets at level $i$ and nodes in level $i+1$, i.e., the node's father-son relationship. However, note that being bipartite does not necessarily means that the graph is level-separable.

\begin{figure}[t]
\centering
\includegraphics[width=0.5\textwidth]{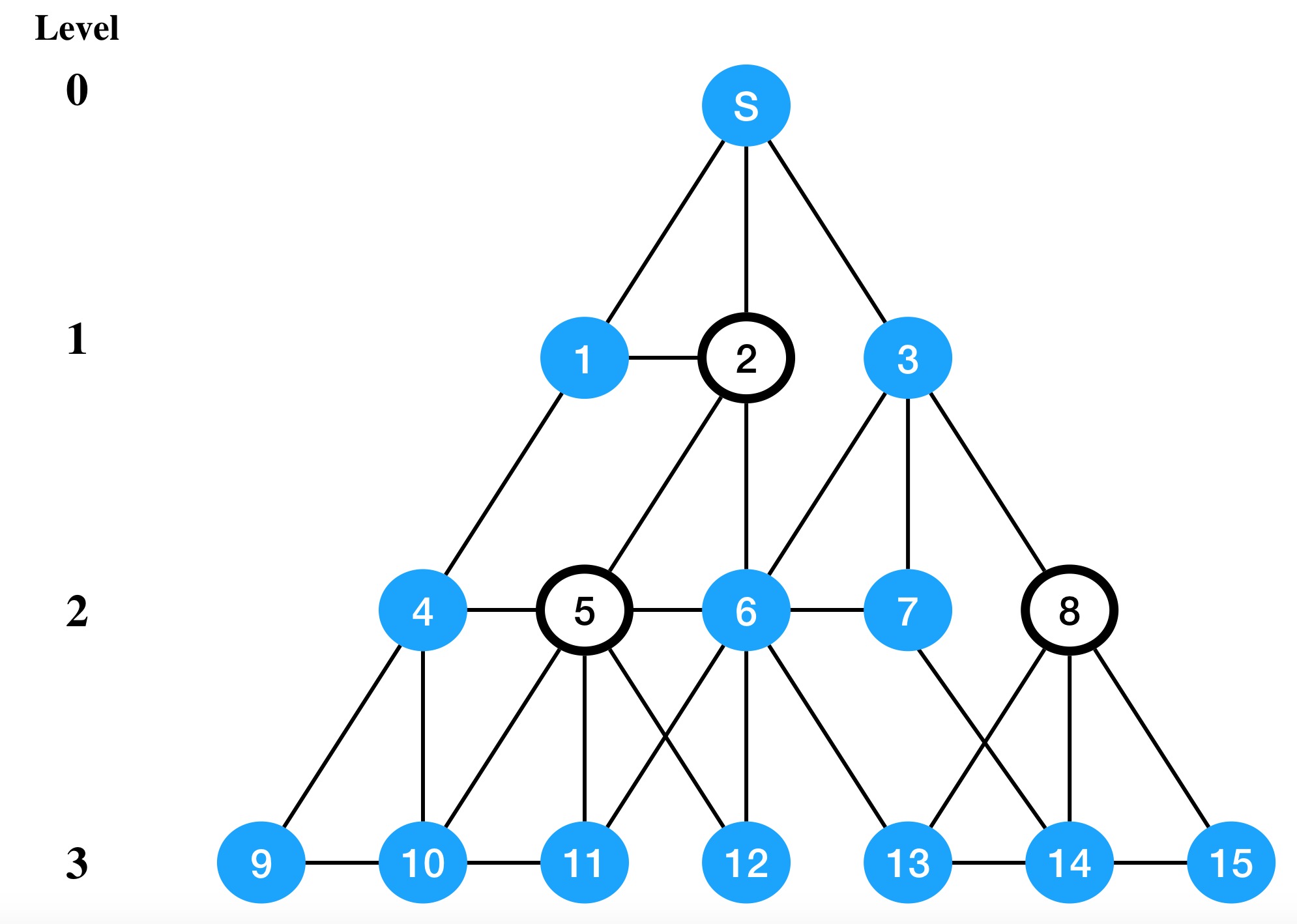}
\caption{Example of a Level-2 separable network, which is not a tree network}
\label{ex2snw}
\end{figure}

Note that a level-separable network is not necessary for being a tree network. However, a tree is a level-separable network. A simple example of a level-separable network is a tree network, where the source node $s$  can be seen as the root of the tree who begins the broadcast. In a tree topology, all non-source nodes have only one parent, i.e. $\forall u \in V-s, \ |P(u)| = 1$. Hence, we can choose $S_{i,1} = S_i$ and $S_{i,2} = \emptyset$. The Level-Separable property is therefore verified. Figure \ref{ex2snw} shows an example of a level-separable network that is not a tree.

Note that studies conducted  in wireless body area networks (e.g. \cite{badreddine2015broadcast}, \cite{badreddine2017peak}, \cite{bu2017total}, \cite{badreddine2017convergecast} and \cite{bu2018ban})   fit our definition of level-separable network.

\section{Broadcast in Level-Separable Network}
In  this section, we propose a 1-bit constant-length labelling broadcast Algorithm  $\beta^{LS}$  detailed in Algorithm \ref{alg1}.  The algorithm  needs $2D$ rounds to terminate, where $D$ is the eccentricity of the broadcast source node.

\subsection{Broadcast with 1-bit Labelling}
Given a level-separable network whose root is the source of the broadcast, we propose Algorithm $\beta^{LS}$ to achieve the wireless broadcast, when a 1-bit labelling scheme $\lambda^{LS}$ is used. Each  node in the network has a 1-bit label, $X_1$. $X_1$ is set to  $1$ or $0$  following the  labelling scheme $\lambda^{LS}$ described below. The idea of the broadcast algorithm is to separate nodes at each level into two independent sets. Nodes in the first set transmit at round $x$ and nodes in the second set transmit at round $x+1$ (the next round), so they will not generate valid collisions\footnote{Note that collisions that occur at a node who has already received the message successfully are not considered as valid collisions.}. 

The broadcast Algorithm $\beta^{LS}$ using the labelling scheme $\lambda^{LS}$ is as follows: the source node sends the message, $\mu$, at round $0$. Nodes at level $1$ receive $\mu$ at the end of round $0$. When nodes with $X_1 = 1$ receive message $\mu$ at round $2i-3$ ($i > 1$) or $2i-2$ ($i>0$), where $i$ is the level, they send $\mu$ at round $2i-1$. When nodes with $X_1 = 0$ receive $\mu$ at round $2i-3$ ($i > 1$) or $2i-2$ ($i > 0$), they send $\mu$ at rounds $2i$. That is, nodes at level $i>0$ will receive $\mu$ from their parents (nodes at level $i-1$) at round $2i-3$ ($i > 1$) or $2i-2$ ($i > 0$), and they will send $\mu$ at round $2i$ or $2i-1$ according to $X_1$. In other words, at each level $i$, nodes take two rounds to propagate $\mu$ to all nodes at level $i+1$. 

\subsection{1-bit Labelling Scheme $\lambda^{LS}$} To achieve collision-free transmission, 1-bit Labelling Scheme $\lambda^{LS}$ $X_1$ of all nodes in $S_{i,1}$ for level $i>0$ is $1$, and $X_1$ of all nodes in $S_{i,2}$ for level $i>0$ is $0$ where $S_{i,1}$ and $S_{i,2}$ are the sets identified in  Definition \ref{def5}. 

\subsection{Correctness of Algorithm $\beta^{LS}$}
\label{subsec32}
In the following, we prove that Algorithm $\beta^{LS}$ is correct. 

\begin{def3}
\label{t1}
 Algorithm $\beta^{LS}$ with 1-bit constant Labelling Scheme $\lambda^{LS}$ implements broadcast in a level-separable network within  $2D$ rounds.
  
\end{def3}
The proof of this theorem is a direct consequence of  Lemmas \ref{lema1}, \ref{lemapp11} and \ref{lemapp22} below.

\begin{def4}
\label{note1}
Note that the 1-bit labelling scheme is optimal for broadcast in a level-separable network. That is, with 0-bit labelling (i.e. without using any labelling) it is possible that some nodes in the network do not receive the broadcasted message due to the collisions since nodes are synchronized and transmit at the same time.
\end{def4}

\begin{def2}
\label{lema1}
Let $G=(V,E)$ be a level-separable network such that each node has a label according to the labelling scheme $\lambda^{LS}$. If nodes with $X_1=1$ at the same level $i \in [1, D-1]$ are the only one to send a message concurrently at round $j$ and on the next round $j+1$ nodes with $X_1=0$ at the same level $i$ are the only one to send a message concurrently, all nodes at level $i+1$ have received the message without collision either at round $j$ or round $j+1$. 
\end{def2}

\textbf{Proof} 
Let $u\in S_{i+1}$. By construction, $u$ has exactly one parent in $S_{i,1}$ or $S_{i,2}$. In the first case, $u$ has received the message without collision at round $j$, and it has received it at round $j+1$ in the second case.

\begin{def2}
\label{lemapp11}
Given a level-separable network whose root is the source node by applying $\beta^{LS}$ and $\lambda^{LS}$ , all nodes in level $i >0$ finish receiving the message $\mu$ at round $2i-2$. 
\end{def2}

\textbf{Proof} 
 We begin from the base case where $i=1$, nodes at level $i=1$ means nodes that are only one hop away from the source node. At round $0$, which is round $2 \times i -2 = 2 \times 1 - 2 = 0$, the source  sends the message. All nodes at level $1$ will receive the message at the end of round $0$. For $i = 2$, as all nodes at level $1$ can receive the message at round $0$,  they will begin to send at round $1$ and round $2$ for nodes in $S_{i,1}$ and $S_{i,2}$, respectively. According to Lemma \ref{lema1}, all nodes received the message without collision at round 2,  which is round $2 \times i -2 = 2 \times 2 - 2 = 2$ and they begin to send the message at round $3$ and $4$. For the general case, we assume that all nodes at level $i, \ i > 2$, finish receiving the message at round $2i-2$. So that nodes begin to send the received message at round $2(i+1)-3$ and $2(i+1)-2$, and nodes at level $i+1$ receive the message at $2(i+1)-3$ and $2(i+1)-2$, that is nodes at level $i+1$ finish receiving the message at round $2(i+1)-2$. 

\begin{def2}
\label{lemapp22}
Given a level-separable network whose root is the source node by applying $\beta^{LS}$ and $\lambda^{LS}$ , the broadcast finishes in $2D$ rounds. 
\end{def2}

\textbf{Proof} 
From Lemma \ref{lemapp11}, nodes having the longest distance to the source will receive the message at round $2D-2$, where $D$ is the  source eccentricity. After receiving the message, these nodes will  send it according to the broadcast algorithm, even though they are already the ending nodes in the network which takes two more rounds. Therefore the broadcast finishes at round $2D$.

%

\begin{algorithm} [t]  \footnotesize
\caption{$\beta^{LS}(\mu)$ executed at each node $v$}
\label{alg1}
\begin{algorithmic} 
\State \%Each node has a variable $sourcemsg$. The source node has this variable initially set to $\mu$, all other nodes have it initially set to $null$. A variable $k$ initially set to $0$ to ensure each node sends $\mu$ only once.
\For{each round $r$ from 0}
\If{$v$ is the source node and $r = 0$}
\State transmit $sourcemsg$
\EndIf
\If{$v$ is not source node and receives $\mu$}
\If{$k = 0$}
\State $sourcemsg \gets \mu$
\If{$r$ is odd number}
\If{$X_1 = 0$}
\State transmit $sourcemsg$ at round $r+3$
\ElsIf{$X_1 = 1$}
\State transmit $sourcemsg$ at round $r+2$
\EndIf
\ElsIf{$r$ is even number}
\If{$X_1 = 0$}
\State transmit $sourcemsg$ at round $r+2$
\ElsIf{$X_1 = 1$}
\State transmit $sourcemsg$ at round $r+1$
\EndIf
\EndIf
\State set $k = 1$
\EndIf
\EndIf
\EndFor
\end{algorithmic}
\end{algorithm}

Consider the execution of the Algorithm $\beta^{LS}$
in a level-separable network with labelling scheme $\lambda^{LS}$, where  nodes in level $i$ have been separated into two sets $S_{i,1}$ and $S_{i,2}$ verifying the level-separable property at level $i$, $\forall i>0$. Nodes in $S_{i,1}$  have $X_1=1$, and nodes in $S_{i, 2}$ have $X_1=0$. The main idea of $\beta^{LS}$ is that,  nodes in each level $i$ separated into two different sets transmit their received messages $\mu$ in different execution rounds to reduce the impact of the collision  at nodes in level $i + 1$. 

According to Algorithm $\beta^{LS}$, the message $\mu$ will be propagated from level to level. Each propagation from a level to the next one takes two execution rounds. In the first round all  nodes in $S_{i, 1}$ send the received message $\mu$. At the end of this round all the nodes that are the sons of nodes in $S_{i, 1}$   receive $\mu$, without collision, see Lemma \ref{lema1}. Therefore sons of nodes in $S_{i, 1}$ contain all the nodes at level $i+1$ who have multi-parents, that means it remains only nodes at level $i + 1$ having only one parent and did not receive $\mu$ yet. In the second  round, all  nodes in $S_{i, 2}$ send $\mu$, and the remaining part of the nodes at level $i+1$ can therefore receive $\mu$ from their unique parent. So that after these two rounds of transmission from level $i$, all the nodes at $i + 1$ will successfully receive the message $\mu$. It takes therefore $2D$ rounds to finish the broadcast. Note that nodes will only send once according to $\beta^{LS}$. Therefore  the algorithm terminates.

\section{Broadcast with ACK in Level-Separable Network}
In this section, we propose a broadcast algorithm with ACK, $\beta^{LS}_{ACK}$, and a Labelling Scheme, $\lambda^{LS}_{ACK}$, for level-separable networks. Our algorithm $\beta^{LS}_{ACK}$ (Algorithm \ref{alg3}) uses only 2-bits labelling and the broadcast finishes  within $2D$ rounds. In our solution,  $ACK$ goes back to the source node in at most $2D$ rounds, where $D$ is the eccentricity of $s$ (the broadcast source node). That means the $ACK$ can be received by the source node at the same round of the broadcast termination.

\subsection{2-bits Labelling Broadcast with ACK}
According to Theorem \ref{t1} the broadcast finishes in a level-separable network within $2D$ rounds where $D$ is the eccentricity of the source node. If the source node has the knowledge of $D$, then it automatically can decide if the broadcast is finished. However, when an $ACK$ is necessary to inform the source node to trigger some additional functions then the source waits for the reception of this message. In order to avoid that $ACK$ takes additional time after the end of the broadcast, we propose to send in advance the $ACK$ message at the halfway of the transmission during the broadcast execution.  Since in a level-separable network, informing nodes from level to level takes exactly $2$ rounds, then $ACK$ also takes 2 rounds to go back one level above. Therefore, when the last node receives $\mu$, the source node  receives $ACK$ at the same round. Interestingly, compared with non-ACK broadcasting, our solution uses one extra bit for labelling and no additional rounds to forwarding $ACK$ back to the source.

\begin{figure*}[t]
\centering
\includegraphics[width=0.7\textwidth]{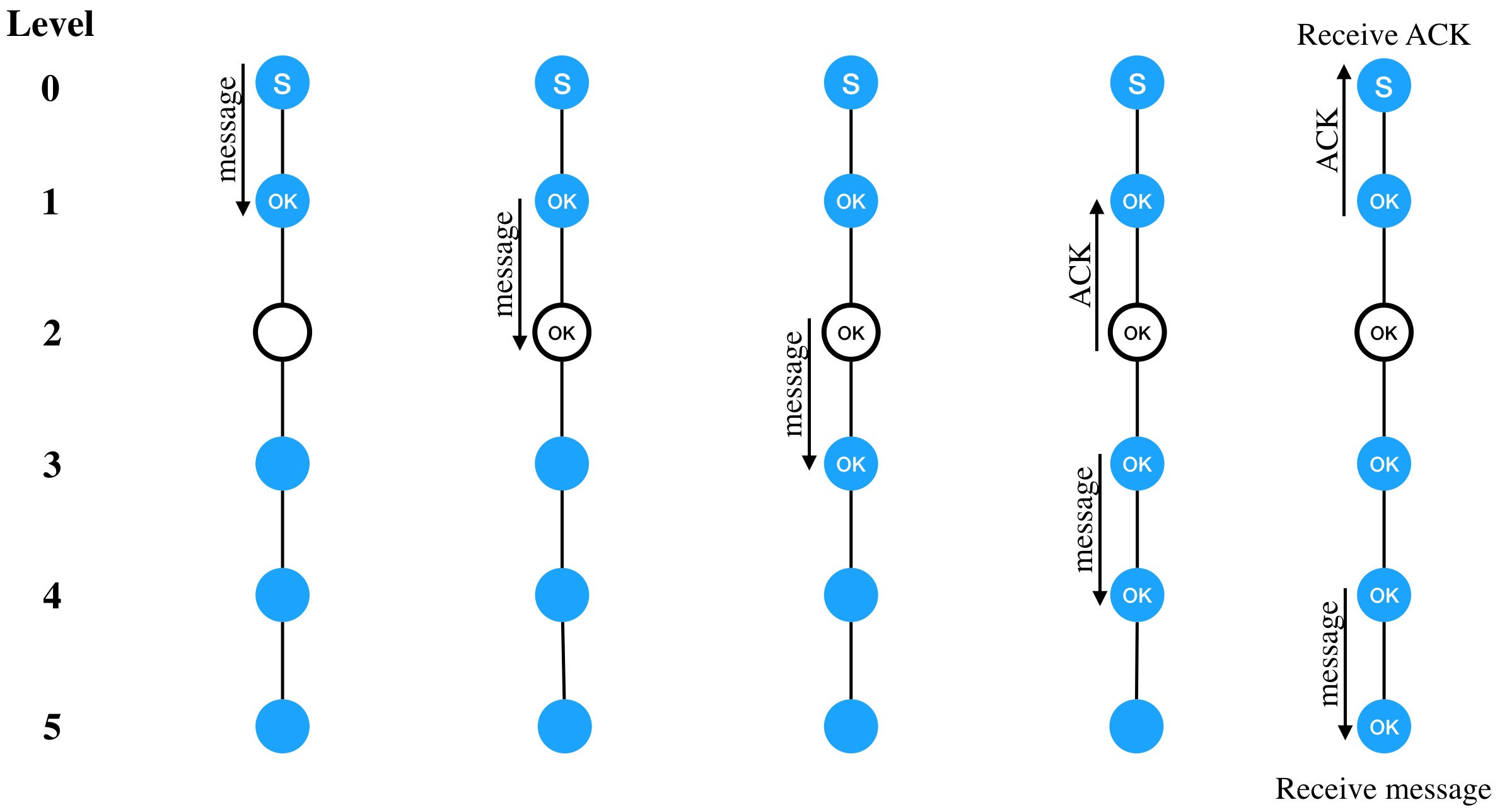}
\caption{Anticipating the ACK  in a level-separable network}
\label{RI1}
\end{figure*}

Figure \ref{RI1} gives the intuition of how to send in advance the $ACK$: the half-way $ACK$ mechanism. In Figure \ref{RI1}, the network is represented in abstract levels to simplify the presentation. Packets flow shown in the figure represent the propagation of messages $\mu$ and $ACK$.
%

\subsection{2-bits Labelling Scheme $\lambda^{LS}_{ACK}$}
We use $\lambda^{LS}$ to set $X_1$ in $\lambda^{LS}_{ACK}$  in order to verify Lemma \ref{lema1}. Let $X_2$ be the second bit of the $\lambda^{LS}_{ACK}$ labelling scheme. $X_2 = 1$ for a set of nodes if they are on the way back  path from a node at level $\lfloor D/2 \rfloor -1$  to the source node, where $D$ is the eccentricity of $s$ and $s$ is the broadcast source. For the other nodes, $X_2 = 0$. In Section \ref{appp3}, we explain why we choose nodes at level $\lfloor D/2 \rfloor -1$ to begin sending the $ACK$.

\begin{def4}
\label{note2}
Note that the 2-bits labelling scheme is optimal to achieve broadcast with acknowledgment in a level-separable network.
From Note \ref{note1} 1-bit is necessary for broadcast without acknowledgment. When an acknowledgment has to be sent back to the source node, at least one additional bit is necessary to indicate  the node to generate the acknowledgment message and send it back to the source node. Without this additional bit no node can decide (unless it uses extra local memory) if it is the last receiving node, and who should send $ACK$ back.
 \end{def4}

\subsection{Correctness of Algorithm $\beta^{LS}_{ACK}$}
\label{appp3}
 
 Theorem \ref{t3} below proves the
  correctness of Algorithm $\beta^{LS}_{ACK}$. 

\begin{def3}
\label{t3}
 Algorithm $\beta^{LS}_{ACK}$ with 2-bits  labelling scheme $\lambda^{LS}_{ACK}$ implements  broadcast in a level-separable network. The broadcast terminates in $2D$ rounds. The ACK message is transmitted back to the source  at round $2(D-1)$, if $D$ is odd or $2D$, if $D$ is even.  
\end{def3}
The proof of the theorem is the direct consequence of  Lemmas \ref{lemapp1},  \ref{lemapp3} and \ref{lemapp4} below.

\begin{def2}
\label{lemapp1}
Given a level-separable network whose root is the source node by applying $\beta_{ACK}^{LS}$ and $\lambda_{ACK}^{LS}$ , nodes in level $i >0$ receive message $\mu$ at round $2i-2$. The broadcast finishes at round $2D$.
\end{def2}
   
%
\textbf{Proof} 
$\beta_{ACK}^{LS}$  follows the same idea as $\beta^{LS}$. The additional $ACK$ transmission will not have any impact according to Lemma \ref{lemapp11} and $\ref{lemapp22}$. Hence the proof follows. 

\begin{def2}
\label{lemapp3}
Given a level-separable network whose root is the source node by applying $\beta_{ACK}^{LS}$ and $\lambda_{ACK}^{LS}$, $ACK$ goes back to the source node at round $2(D-1)$, if $D$ is odd; or $2D$, if $D$ is even.
\end{def2}

\textbf{Proof} 
When $D$ is odd, $ACK$ and $\mu$ will begin to be sent to source and to the ending nodes from levels $l_{ACK}$ and $l_{MSG}$, respectively. The distances from levels $l_{ACK}$ back to the source are the same as that from $l_{MSG}$ to the ending nodes. $ACK$ arrives at the source at the same round as $\mu$ arrives at the ending nodes. According to Lemma \ref{lemapp1},  this is round $2(D-1)$. When $D$ is even $ACK$ needs to go one level farther compared with $\mu$. Therefore, it takes two extra rounds when $D$ is even. Therefore, when $D$ is even the $ACK$ message goes back to the source node in $2D$ rounds.

\begin{def2}
\label{lemapp4}
Given a Level-Separable Network whose root is the source node by applying $\beta_{ACK}^{LS}$ and $\lambda_{ACK}^{LS}$ , the algorithm finishes within $2D$ rounds.
\end{def2}

\begin{algorithm} [t]  \footnotesize
\caption{$\beta^{LS}_{ACK}(\mu)$ executed at each node $v$}
\label{alg3}
\begin{algorithmic}
\State \%Each node has a variable $sourcemsg$. The source node has this variable initially set to $\mu$, all other nodes have it initially set to $null$. A variable $k$ and $k_{ack}$ initially set to $0$ to ensure each node send $\mu$ only once. 
\For{each round $r$ from 0}
\If{$v$ is source node and $r$ = 0}
\State transmit $sourcemsg$
\EndIf
\If{$v$ is not source node and received $\mu$}
\State $sourcemsg \gets \mu$
\If{$k = 0$}
\If{$r$ is odd number}
\If{$X_1 = 0$}
\State transmit $sourcemsg$ at round $r+3$
\If{$X_2 = 1$}
\State  transmit "pACK" at round $r+4$
\If{$v$ does not received "pACK" at $r+6$}
\State  transmit "ACK" at round $r+6$, set $k_{ack} = 1$
\EndIf 
\EndIf 
\ElsIf{$X_1 = 1$}
\State transmit $sourcemsg$ at round $r+2$
\If{$X_2 = 1$}
\State  transmit "pACK" at round $r+4$
\If{$v$ has not received "pACK" at $r+6$}
\State  transmit "ACK" at round $r+6$, set $k_{ack} = 1$
\EndIf 
\EndIf 
\EndIf
\ElsIf{$r$ is even number}
\If{$X_1 = 0$}
\State transmit $sourcemsg$ at round $r+2$
\If{$X_2 = 1$}
\State  transmit "pACK" at round $r+3$
\If{$v$ has not received "pACK" at $r+5$}
\State  transmit "ACK" at round $r+5$, set $k_{ack} = 1$
\EndIf 
\EndIf 
\ElsIf{$X_1 = 1$}
\State transmit $sourcemsg$ at round $r+1$
\If{$X_2 = 1$}
\State  transmit "pACK" at round $r+3$
\If{$v$ has not received "pACK" at $r+5$}
\State  transmit "ACK" at round $r+5$, set $k_{ack} = 1$
\EndIf 
\EndIf 
\EndIf
\EndIf
\State set $k = 1$
\EndIf
\EndIf
\If{$v$ is not source node and received $ACK$}
\If{$X_2 = 1$ and $k_{ack}$ = 0}
\State transmit $ACK$ at round $r+2$
\State set $k_{ack} = 1$
\EndIf
\EndIf
\EndFor
\end{algorithmic}
\end{algorithm}


\textbf{Proof} 
The idea of the correctness proof is as follows. Consider a level-separable network  with the labelling scheme $\lambda^{LS}_{ACK}$, where all  nodes in level $i$ have been separated into two sets $S_{i,1}$ and $S_{i,2}$. Nodes in $S_{i,1}$ have $X_1=1$, and nodes in $S_{i, 2}$ have $X_1=0$. A way back path is marked with $X_2 = 1$ between source $s$ and an arbitrary node at level $\lfloor D/2 \rfloor -1$, where $D$ is the eccentricity of $s$ , i.e., we only mark the way back  path from the half-way level $\lfloor D/2 \rfloor -1$ of the network in this case.

The idea is that when the message $\mu$ propagates to the half-way level of the network, a node at that level will begin $ACK$   transmission processing, so that when the $\mu$ reaches to the ending node(s) at level $D$, $ACK$ reaches the source $s$ at (almost) the same round. As nodes cannot decide if they are the ones at the half-way of the network who should generate and send $ACK$, we use a \emph{Waiting Period} and an extra $pACK$ message.

According to the $\beta^{LS}_{ACK}$, when a node with $X_2 = 1$, receives $\mu$ and finishes the $\mu$ retransmission, it cannot decide its position in the way back path. Therefore, it sends a $pACK$ and begins to wait for $pACK$ message sent to him in the following rounds. When a node with $X_2 = 1$ receives a $pACK$ within the $Waiting Period$, that means it is not the ending node, because there is another node with $X_2 = 1$ that received $\mu$ and sent $pACK$ to him. When a node with $X_2 = 1$ does not receive any $pACK$ within its $Waiting Period$, this means no node in the next level has $X_2 = 1$, i.e., it is the half-way ending node, so it generates and sends the $ACK$. All the nodes with $X_2 = 1$ will forward $ACK$ from the ending node to the source $s$ according to the marked way back path. In the $\beta^{LS}_{ACK}$, the $Waiting Period$ is delayed two rounds after a node sends $pACK$ to avoid the collision between $pACK/ACK$ and $\mu$. 

A node with $X_2 = 1$ that receives $\mu$ at round $x$, transmits $\mu$ at round $x+2$, then it sends $pACK$ to its parents at round $x+4$, then it waits a \emph{Waiting Period} until round $x+6$. If it doesn't receive another $pACK$, then it sends $ACK$ at round $x+8$. That means, for the half-way ending node, it needs to wait for 6 rounds to begin sending $ACK$. What we want for this half-way mechanism is that the source node can receive $ACK$ as fast as possible, after the broadcast finishes. When $D$ (the eccentricity of the broadcast source $s$) is odd, then if we chose the node at level $\lfloor D/2 \rfloor -1$ as the half-way ending node, then the $ACK$ can be received by the source node at the same round as the end of the broadcast. Because after waiting for 6 rounds at level $\lfloor D/2 \rfloor -1$, $\mu$ has already been transmitted to level $\lfloor D/2 \rfloor -1 + 3 = \lfloor D/2 \rfloor +2$. The distance from node sending $ACK$ to source node is $d(s, \lfloor D/2 \rfloor -1) = \lfloor D/2 \rfloor -1$; the distance from node sending $\mu$ to nodes at level $D$ is also $d(\lfloor D/2 \rfloor +2, D) =  \lfloor D/2 \rfloor -1$. When $D$ is even, if we chose the node at level $\lfloor D/2 \rfloor -1$ as the half-way ending node, then the $ACK$ can be received by the source node only two rounds after the round of the ending of the broadcast.

 Therefore it takes $2D$ rounds to finish the broadcast and $ACK$ can be transmitted back to the source node at round $2(D-1)$ or round $2D$. Note that nodes will only send (both for $\mu$ and $ACK$) once according to $\beta^{LS}_{ACK}$. Therefore the algorithm terminates.

\section{Hardness of  level separation}
\label{NP}
It should be noted that checking that a separation is a Level-Separation is  polynomial: it is sufficient to check that  for each node $u$, $|P(u)\cap S_{l(u),1}|=1\vee |P(u)\cap S_{l(u),2}|=1$. In this section we will prove that determining if a graph has the level-separable property is NP-Hard.  To do so, we will reduce 1-IN-3-SAT~\cite{johnson1979computers} to the level separable problem.
1-IN-3-SAT is a NP-Complete variant of the usual NP-complete problem 3-SAT, where exactly a single literal in each clause must be true. 
As input, we have a list of variables $X=\{x_1,\ldots,x_k\}$ and a formula $\phi$ which is a conjunction of clauses $c_1,\ldots,c_l$ that are each composed of exactly 3 literals of the form $x_i$ or $\overline{x_i}$.
The goal is to find an assignation for the variables $A:X\to\{\top,\bot\}$ such that, for every clause $c_i$, exactly one variable is satisfied (i.e. has the assignation $\top$ if it appears positively, $\bot$ if it appears negatively).

\begin{def3}
Determining if a graph with a source has the Level-Separable property is NP-complete.
\end{def3}

\textbf{Proof} 
Let $(X,\phi)$ be an instance of 1-IN-3-SAT. We will build $G=(V,E)$ such that $V=\{s\}\cup S_1\cup S_2$, $S_1$ being the neighborhood of $s$, and $S_2$ all the other nodes, that will actually be at distance 2 from $s$. We have:
\begin{itemize}
\item $S_1=\{u_{n_a},u_{n_b},u_{y}\}\cup\{u_{y_i},u_{n_i}\}_{i\le k}$.
\item $S_2=\{v_{a},v_{b}\}\cup\{v_{x_i}\}_{i\le k}\cup\{v_{c_j}\}_{j\le l}$.
\item $\{s\}\times S_1\subset E$, $\{(u_{n_a},v_a), (u_{n_b}, v_b), (u_{y}, v_a),(u_{y}, v_b)\}\subset E$.
\item $\forall i\le k$, we have $\{(u_{y_i}, v_{x_{i}}),(u_{n_i}, v_{x_{i}})\}\subset E$. If $x_i\in c_j$, then we have $(u_{y_{i}},v_{c_j})\in E$.  If $\overline{x_i}\in c_j$, then we have $(u_{n_{i}},v_{c_j})\in E$.
\item $\forall j\le l$, we have $\{(u_{n_a}, v_{c_{j}}),(u_{n_b}, v_{c_{j}})\}\subset E$.
\end{itemize}

\begin{figure}[t]
\centering
\includegraphics[width=0.5\textwidth]{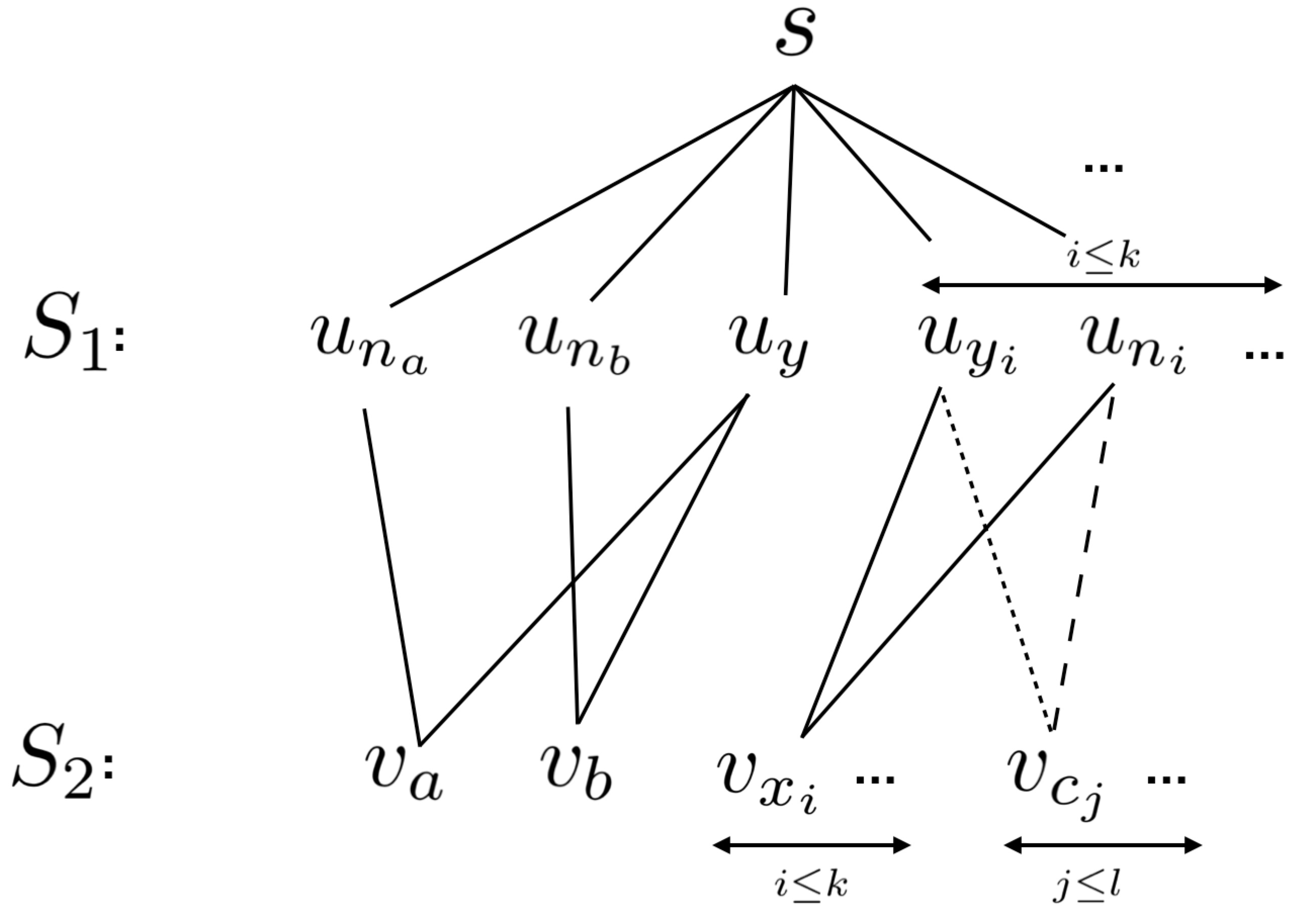}
\caption{Solid lines represent the edges that always exist. Dense dotted lines represent the edges that exist if $x_i\in c_j$. Loose dotted lines represent the edges that exist if $\overline{x_i}\in c_j$}
\label{3sat}
\end{figure}

An abstract graph can be seen in Figure \ref{3sat} corresponding to the description above.

Let's suppose that we have a solution $S_{1,1},S_{1,2}$ to the problem (any partition of $S_2$ works, as there are the farthest nodes from $s$). We will call $Y\in\{1,2\}$ the index of the node $u_y$, and $N=3-Y$ the index that is different from $Y$. Here below a list of observations:
\begin{enumerate}
\item If a node in $S_2$ has exactly two parents, then the index of its parents must be different.
\item $u_{n_a}$ (resp. $u_{n_b}$) must have index $N$, as $v_a$ (resp. $v_b$) is only connected to it and to $u_y$.
\item $\forall i\le k$, $u_{y_i}$ and $u_{n_i}$ have different indexes, as they are the only parents of $v_{x_i}$.
\item $\forall j\le l$, $v_{c_j}$ has exactly one parent of index $Y$, as it has at least two parents of index $N$: $u_{n_a}$ and $u_{n_b}$.
\end{enumerate}

A solution for the corresponding 1-IN-3-SAT instance is to choose, for each variable $x_i$ such that $u_{y_i}$ has index $Y$, valuation $\top$, and $\bot$ for the others. Let $c_j$ be a clause. The node $v_{c_j}$ has exactly one parent of index $Y$ among the ones corresponding to variables. If it is a node of the form $u_{y_i}$, then $x_i$ appears positively in $c_j$ (otherwise, it is of the form $u_{n_i}$ and $x_i$ appears negatively in $c_j$). Let be another node corresponding to a variable connected to $v_{c_j}$. Its index must be $N$, and it appears positively in $c_j$ iff the node is some $u_{y_i}$ iff we chose $\bot$ for $x_i$. Hence, $c_j$ has exactly one variable satisfied.\\

Reversely, let's suppose that we have an assignation $A$ to the 1-IN-3-SAT instance. We choose $S_{1,1}=\{u_y\}\cup\{u_{y_i}: A(x_i)=\top\}\cup\{u_{n_i}: A(x_i)=\bot\}$ and $S_{1,2}=S_1\setminus S_{1,1}$. Let's prove that each node in $S_2$ has exactly one parent in $S_{1,1}$. For $v_{a}$ and $v_{b}$, it is $u_y$. For a node $v_{c_j}$, we know that exactly one variable of $c_j$ is satisfied. Its corresponding node is in $S_{1,1}$ by construction, and the corresponding node of the two other variables are in $S_{1,2}$ by construction. As $v_{n_a}$ and $v_{n_b}$ are also in $S_{1,2}$, this concludes the proof.


%
%
%
%

\section{Conclusion}
We proposed solutions for implementing  broadcast   in wireless networks when the broadcast is helped by a labelling scheme.  We studied broadcast  without acknowledgment (i.e. the initiator of the broadcast  is not notified at the end of the broadcast) and broadcast with acknowledgment. We first improved in terms of memory complexity the scheme proposed in   \cite{ellen2019constant} for arbitrary networks. Then we propose an optimal acknowledgment-free broadcast strategy using only 1-bit labelling  and a broadcast with acknowledgment using a 2-bits labelling  
in  level 2-separable networks. The complexity of both  algorithms is $2D$ where $D$ is the eccentricity of the  broadcast initiator.  Level 2-separable networks have a practical interest in the large literature of WBAN.

In Section \ref{NP}, we proved that the verification of the level-separable property can be done in polynomial time while determining if a graph has the level separable property is  NP-hard. This result may be considered as 
a serious break in exploiting the level separable property in  labelling-based algorithms.
  However, in the case of small scale networks such as WBAN,  polynomial algorithms may be of   practical interest. 
For the case of large scale networks, since  the verification of the level-separable property is NP-hard, we recommend to  exploit  $MIMO$ antenna technology \cite{haimovich2007mimo, li2007mimo} (wireless devices  having the capability to focus the wireless transmission on several dedicated directions).  
Thanks to  this technology the connections from a node to  several of its neighbours can be disabled. 
This simple mechanism can help in constructing networks with built-in level separable property according to the description in Section \ref{defLVS}. In this case, our algorithms are the best to date for labelling-based broadcast.
 
 Independent of the practical interest of our work, an interesting theoretical research direction is opened by our study: the generalization of our results to  level k-separable networks. In this framework, it would be interesting to find  optimal separations for a  graph and the 
 tradeoff between the time and the bit complexity  of broadcast in level k-separable networks. 
\bibliographystyle{plain}
\bibliography{sample-dmtcs,bib}

\end{document}